\documentclass[12pt]{article}
\usepackage{jcappub}

\usepackage[utf8]{inputenc}
\usepackage{subfig}

\usepackage{amsmath, amssymb}
\usepackage{graphicx}

\usepackage{color}

\newcommand{\ud}{\mathrm{d}}

\newcommand{\Mpl}{M_{\mathrm{pl}}}

\newcommand{\dotsig}{\dot\sigma}
\newcommand{\ddotsig}{\ddot \sigma}

\newcommand{\J}{\text{J}}
\newcommand{\pn}[1]{\left(#1\right)}

\title{Double inflation via non-minimally coupled spectator}

\author[a]{Mio Kubota,}
\affiliation[a]{Department of Physics, Ochanomizu University, Tokyo 112-8610, Japan}

\author[b]{Kin-ya Oda,}
\affiliation[b]{Department of Mathematics, Tokyo Woman's Christian University, Tokyo 167-8585, Japan}

\author[c]{Stanislav Rusak,}
\affiliation[c]{Departamento de F\'{i}sica Te\'{o}rica and Instituto de F\'{i}sica de Part\'{i}culas y del Cosmos IPARCOS,
Universidad Complutense de Madrid, E-28040 Madrid, Spain}

\author[d]{and Tomo~Takahashi}
\affiliation[d]{Department of Physics, Saga University, Saga 840-8502, Japan}

\abstract{
We argue that double inflation may occur when a spectator field is non-minimally coupled to gravity. As a concrete example, we study a two-field inflationary model where the initial spectator field is non-minimally coupled to gravity while the initial inflaton field is minimally coupled. The non-minimal coupling results in the growth of the spectator field which, in turn, drives the second stage of inflation in a significant region of parameter space. The isocurvature fluctuations originating from the spectator field source adiabatic ones, and hence the spectator non-minimal coupling can modify the inflationary predictions for the spectral index and the tensor-to-scalar ratio even though the initial inflaton field is minimally coupled to gravity. We explicitly show that quadratic chaotic inflation can become viable by the introduction of the spectator non-minimal coupling.
} 

\begin{document}

\rightline{OCHA-PP-370}

\maketitle

\section{Introduction}  
\label{sec:intro}

The Universe is believed to have experienced a period of nearly exponential expansion, dubbed inflation, in its very early stage.  This inflationary period not only helps address the flatness and horizon problems but also generically provides a mechanism to generate primordial density fluctuations necessary for structure formation. A simple possibility is that a single scalar field, called inflaton, is responsible for both driving the inflationary expansion and also generating the primordial fluctuations. Since the nature of the primordial fluctuations depends on models of the inflaton, cosmological observations such as the cosmic microwave background (CMB) can probe these models, and indeed, data from the Planck satellite in combination with other observations, especially CMB B-mode polarization experiments such as BICEP/Keck, provide constraints on these models~\cite{Planck:2018jri,BICEP:2021xfz}.

Though having only a single field responsible for inflation is attractive, the simplest inflationary potentials are now severely constrained by cosmological data and many have been ruled out.  Therefore multi-field models have also been extensively discussed in the literature. These include cases where the inflaton is only responsible for the inflationary expansion while another scalar field, whose energy density is negligible during inflation,  and irrelevant to the inflationary dynamics (referred to as a spectator field),  is responsible for generating primordial fluctuations. Examples of such models include the curvaton model \cite{Enqvist:2001zp,Lyth:2001nq,Moroi:2001ct} and modulated reheating scenarios \cite{Dvali:2003em,Kofman:2003nx}. In high energy theories such as string theory, the existence of multiple scalar fields is generic, which also motivates a multi-field inflation model. In addition, when a spectator field is responsible for primordial fluctuations, the predictions for primordial power spectra are modified and, in some cases, even if the model is excluded by the data as a single-field inflation model, it can be resurrected in the framework of spectator field models; see, e.g., \cite{Langlois:2004nn,Moroi:2005kz,Moroi:2005np,Ichikawa:2008iq,Enqvist:2013paa,Vennin:2015vfa,Fujita:2014iaa} for the curvaton model and \cite{Ichikawa:2008ne} for the modulated reheating scenario.

Another extension of a single-field inflation model is to introduce a non-minimal coupling to gravity. A well-known example is the Higgs inflation~\cite{Salopek:1988qh,Bezrukov:2007ep} where the inflaton is identified with the Standard Model Higgs field with non-minimal coupling to gravity.\footnote{
There have been early works along this line. See \cite{Spokoiny:1984bd,Futamase:1987ua,Fakir:1990eg,Komatsu:1999mt} for such early studies.
}
The potential for the Higgs field is quartic during inflation in the pure Standard Model, and therefore the pure Standard Model Higgs inflation is excluded by observational data. However, by introducing a non-minimal coupling, the model becomes consistent with cosmological observations in terms of the spectral index and the tensor-to-scalar ratio (the suppression of the tensor-to-scalar ratio has already been discussed in \cite{Komatsu:1999mt}). Inflationary models with non-minimal coupling have become an extensively studied research area as they generally predict modified inflationary observables; see, e.g., \cite{Linde:2011nh,Boubekeur:2015xza,Tenkanen:2017jih,Ferreira:2018nav,Artymowski:2018ewb,Antoniadis:2018yfq,Takahashi:2018brt,Jinno:2018jei,Shokri:2019rfi,Takahashi:2020car,Reyimuaji:2020goi,Cheong:2021kyc,Kodama:2021yrm}.  In addition, non-minimal couplings generically arise radiatively in quantum field theory in curved space-time \cite{Birrell:1982ix,Parker:2009uva} even if they do not exist at tree level. Therefore inflation models with non-minimal coupling to gravity are well motivated theoretically. 

Given that both multi-field inflationary models and non-minimal couplings are well motivated, it is tempting to consider multi-field models with non-minimal couplings; see, e.g., \cite{Kaiser:2010yu,White:2012ya,Kaiser:2012ak,Greenwood:2012aj,Kaiser:2013sna,White:2013ufa,Schutz:2013fua,Karamitsos:2017elm,Liu:2020zzv,Modak:2020fij,Lee:2021rzy} for works along this line. In this paper, we consider a multi-field model in which there exist two scalar fields: a minimally coupled field $\Phi$ and non-minimally coupled one $S$, with $\Phi$ being the inflaton and $S$ a spectator initially.  As will be shown in this paper, due to the spectator non-minimal coupling, its amplitude can grow during the inflationary phase driven by the $\Phi$ field until eventually the $S$ field acquires super-Planckian field values, which, in turn,  drives a second inflationary phase after the end of the first inflationary epoch driven by the $\Phi$ field.\footnote{
Even in minimally coupled models, when the spectator field is assumed to have a super-Planckian amplitude, the second inflationary phase can be driven by the spectator field (these kinds of models are sometimes called ``inflating curvaton models")~\cite{Langlois:2004nn,Moroi:2005kz,Ichikawa:2008iq,Dimopoulos:2011gb,Enqvist:2019jkb}.
}

We also discuss primordial fluctuations in such models. Since isocurvature fluctuations generated by the spectator field can source the adiabatic ones,  the existence of the spectator non-minimal coupling can also modify primordial power spectra from its minimally coupled counterpart.  In particular, the predictions for the spectral index $n_s$ and the tensor-to-scalar ratio $r$ are modified. This issue has already been discussed for single-field non-minimally coupled models (see, e.g., \cite{Linde:2011nh,Boubekeur:2015xza,Tenkanen:2017jih,Ferreira:2018nav,Antoniadis:2018yfq,Takahashi:2018brt,Shokri:2019rfi,Takahashi:2020car,Reyimuaji:2020goi,Cheong:2021kyc,Kodama:2021yrm}). We show in this paper that the spectator non-minimal coupling can also relax the constraints on $n_s$ and $r$ so that even inflation models that are excluded in the single-field minimally coupled framework, such as the quadratic chaotic inflation, can become viable due to the existence of the spectator non-minimal coupling.

The structure of this paper is the following. In the next section, we describe the setup for the multi-field inflation model with the non-minimal coupling we consider in this paper. Then in Section~\ref{sec:background}, we discuss the background evolution,  particularly focusing on the growth of the spectator field during the initial inflationary phase driven by $\Phi$. We also argue that in our setup, a second inflationary phase is somewhat generic due to this growth originating from the existence of the spectator non-minimal coupling. In Section~\ref{sec:perturbation}, we discuss primordial scalar and tensor power spectra in the model and how they are modified by changing the strength of the non-minimal coupling and the masses of the fields.  We also investigate how the inflationary predictions are modified to relax the constraints on inflation models. The final section is devoted to conclusions and discussion.

\section{Setup \label{sec:setup}}

We consider two scalar fields $\Phi$ and $S$ that play the role of inflaton and spectator, respectively, at the early stages of inflation.
For simplicity, we assume that $\Phi$ is minimally coupled to gravity, while $S$  non-minimally at the leading order such that the action in the Jordan frame is given by 
\begin{equation}
\label{eq:S_jordan}
	S_{\rm J} 
	= \int \ud^4x \sqrt{-g}\left(\frac{1}{2} M_{\rm pl}^2 f(S)  R
	- \frac{1}{2} \partial_{\mu}\Phi \, \partial^{\mu}\Phi 
	- \frac{1}{2} \partial_{\mu} S \, \partial^{\mu} S
	- V_\J (\Phi, S) \right) ,
\end{equation}
where $M_{\rm pl}=1/\sqrt{8\pi G}=2.4\times10^{18}\,\text{GeV}$ is the reduced Planck mass and $f(S)$ characterizes a non-minimal coupling between $S$ and $R$:
\begin{equation}
\label{eq:f_def}
f(S) =  1+  \xi_S \left( \frac{S}{M_{\rm pl}} \right)^2,
\end{equation}
with $\xi_S$ being the non-minimal coupling constant,  and $V_\J$ is the Jordan frame potential.  As a simplest non-trivial example, we assume that the potential takes the following form:
\begin{equation}
\label{eq:F}
V_\J(\Phi, S) =  \frac12 m_\Phi^2 \Phi^2 + \frac12 m_S^2 S^2.
\end{equation}

To follow the evolution of the scalar fields and the cosmic expansion,  we work in the Einstein frame.  For a system with multiple scalar fields, one can write the action in the Einstein frame as\footnote{
For the formalism of a general multi-field inflation, see e.g., \cite{Langlois:2008mn,Kaiser:2010yu,Kaiser:2012ak}. We follow the notation of \cite{Langlois:2008mn}.
}
\begin{equation}
\label{eq:S_einstein}
 S = \int \ud^4 x \sqrt{-g}\left[ \frac12 \Mpl^2 R - \frac{1}{2}G_{IJ}\nabla_{\mu}\varphi^I\nabla^{\mu}\varphi^J - V(\varphi^I)\right], 
\end{equation}
where the indices run for $I=\Phi,S$ with the abbreviation $\pn{\varphi^\Phi,\varphi^S}=\pn{\Phi,S}$ for the field space. Here, $G_{IJ}$ is the field space metric given by
\begin{equation}
\label{eq:G_IJ}
G_{\Phi\Phi} = f^{-1} , 
 \qquad 
G_{SS} = f^{-1} +  \frac{3}{2}\frac{f_S f_S}{f^2}  , 
\qquad
G_{\Phi S} = G_{S\Phi} = 0,
\end{equation}
with $f_S = \partial  f / \partial S = 2 \xi_S S / \Mpl^2$, and $V(\varphi^I)$ is the Einstein-frame potential related to the Jordan-frame counterpart as 
\begin{equation}
V(\varphi^I) = \frac{V_\J(\varphi^I)}{f^2} .
\end{equation}

\section{Background Evolution \label{sec:background}}

Let us first discuss the background evolution of the fields. We first focus on the case where the non-minimal coupling is small and therefore the field space metric component of the spectator can be approximated as $G_{SS} \simeq f^{-1}$. We can then introduce a canonically normalized spectator field\footnote{
The situation for the spectator is analogous to Higgs inflation in Palatini gravity~\cite{Bauer:2008zj,Rasanen:2017ivk,Jarv:2017azx,Jinno:2019und}.
}
\begin{equation}
\chi \equiv \frac{M_{\rm pl}}{\sqrt{\xi_S}} \,\,  {\rm arcsinh} \left( \sqrt{\xi_S} \frac{S}{M_{\rm pl}} \right)
\end{equation} 
to aid the intuitive understanding of the dynamics. Though this rescaling introduces some kinetic couplings between the spectator and inflaton fields, in the double inflation regime which we are interested in, such couplings will be unimportant. This is due firstly because of slow-roll dynamics and secondly due to relative smallness of the term $\xi_S S^2$ necessary to achieve desirable duration of inflation as we shall discuss.

The equation of motion for the inflaton is then given by

\begin{equation}
 \ddot\Phi + \left[3H - \frac{2\sqrt{\xi_S}}{\Mpl}\tanh\left(\sqrt{\xi_S}\Mpl^{-1}\chi\right)\dot\chi\right]\dot\Phi + \frac{m_\Phi^2\Phi}{\cosh^2\left(\sqrt{\xi_S}\Mpl^{-1}\chi\right)} = 0.
\end{equation}
The presence of the spectator acts to scale down the effective mass of the inflaton and to introduce an extra anti-friction term. For slow-roll dynamics and moderate values of the spectator amplitude, however, these changes are negligible and the dynamics is virtually that of the usual quadratic chaotic inflation. The equation of motion for the canonical spectator, on the other hand, is

\begin{equation}
  \ddot\chi +3H\dot\chi-\frac{2\sqrt{\xi_S}}{\Mpl}\frac{\sinh(\sqrt{\xi_S}\Mpl^{-1}\chi)}{\cosh^5(\sqrt{\xi_S}\Mpl^{-1}\chi)}\left[m_\Phi^2\Phi^2-\frac{m_S^2\Mpl^2}{2\xi_S}\left(1-\sinh^2(\sqrt{\xi_S}\Mpl^{-1}\chi)\right)\right]=0,
\end{equation}
where we have neglected the term proportional to $\dot\Phi^2$ as it is negligible during slow-roll.

We can then deduce the dynamics as follows. Initially, the inflaton field dominates and induces a force that amplifies the spectator field. Then, as the amplitude of $\Phi$ decreases past the threshold

\begin{equation}
 \Phi_\mathrm{turn} = \frac{\Mpl m_S}{\sqrt{2\xi_S} m_\Phi},
\end{equation}
the force reverses and the spectator is driven back toward zero as long as its amplitude has not exceeded the critical value

\begin{equation}
 \chi_\mathrm{runaway} = \frac{\Mpl}{\sqrt{\xi_S}}\operatorname{arcsinh}(1) \qquad \mathrm{or} \qquad S_\mathrm{runaway} = \frac{\Mpl}{\sqrt{\xi_S}}.
\end{equation}
If this value is exceeded, the spectator is amplified indefinitely.

\begin{figure}
\begin{center}
\includegraphics[width=.7\textwidth]{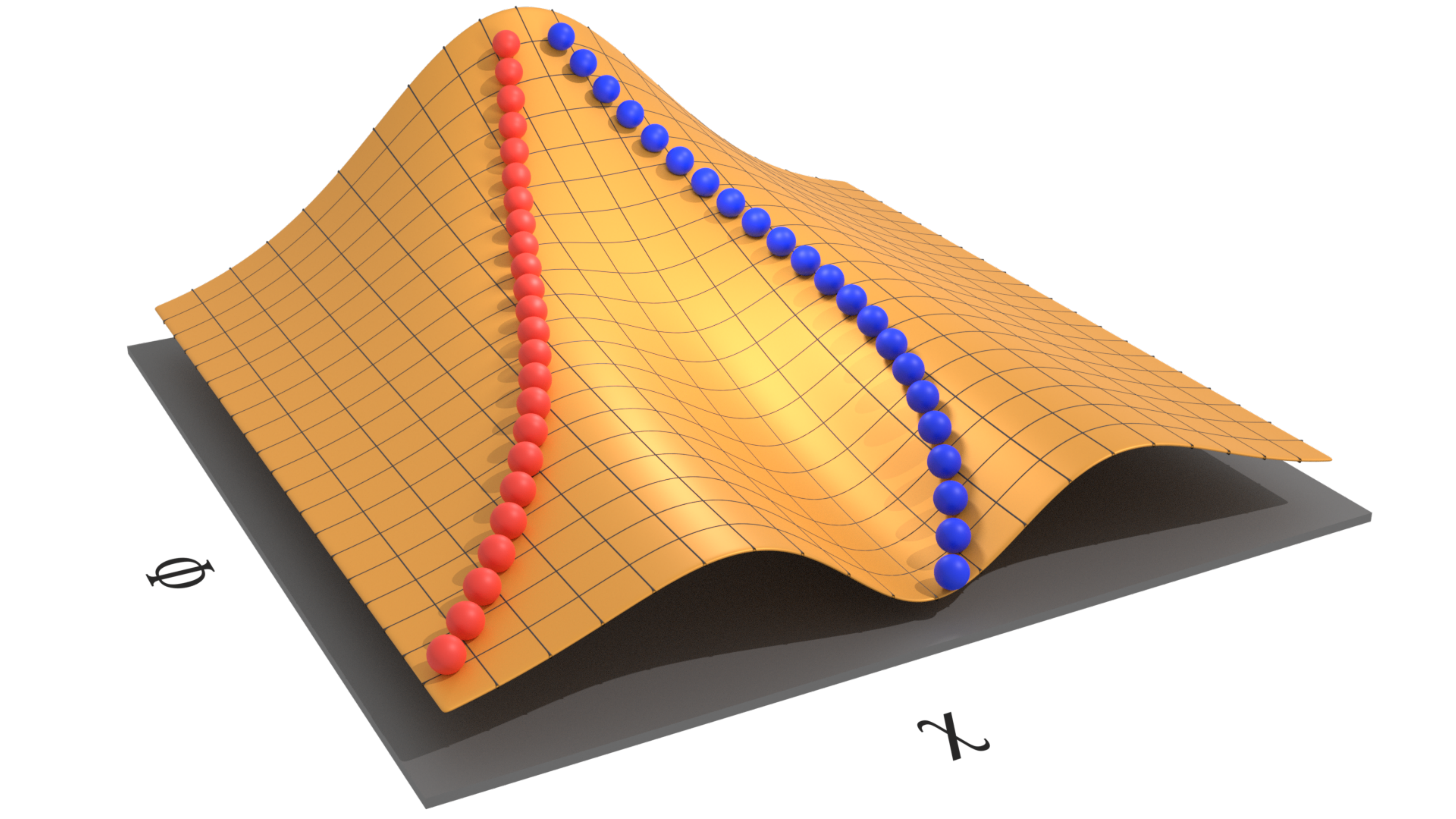}
\caption{The effective potential for the initial inflaton $\Phi$ and the initial rescaled spectator $\chi$. In the $\chi$ direction, the potential switches from being convex at large $\Phi$ to being concave around the origin of $\chi$ at small $\Phi$, causing $\chi$ to first be amplified and then roll back (blue trajectory). If $\chi$ amplitude grows past the ridge of the potential, however, it continues to be amplified indefinitely resulting in runaway behaviour (red trajectory).}
\label{fig:V_eff}
\end{center}
\end{figure}

When the spectator is amplified to a large value, it comes to dominate the energy density, driving the second stage of inflation while it rolls back to zero. Essentially the physics in the regime of interest is described by the effective potential
\begin{equation}
\label{eq:V_eff}
V_{\rm eff} 
= 
\frac{
\xi_S m_\Phi^2  \Phi^2 + m_S^2 M_{\rm pl}^2  ~{\rm sinh}^2 (\sqrt{\xi_S} \chi / M_{\rm pl})
}{
2\xi_S ~ {\rm cosh}^4  (\sqrt{\xi_S} \chi / M_{\rm pl})
},
\end{equation}
which is depicted in Fig.~\ref{fig:V_eff}: When $\Phi$ is large (upper left direction), $V_\text{eff}$ is convex in the direction of $\chi$, causing it to be amplified. At smaller values of $\Phi$ (lower right direction), $V_\text{eff}$ becomes concave in the $\chi$ direction, causing it to roll back, and drives the second inflationary stage.

\subsection{Numerical analysis}

Having gained an analytical understanding of the dynamics we proceed to numerically solve the full equations of motion 
\begin{equation}
 \mathcal{D}_t\dot\varphi^I + 3 H\dot\varphi + G^{IJ}V_{,J} = 0
\end{equation}
which is derived from the action~\eqref{eq:S_einstein}. Here $\mathcal{D}_t$ is the field-space-covariant time derivative: $\mathcal D_t\dot\phi^I\ = \ddot \phi^I + \Gamma^I{}_{JK}\dot\phi^J\dot\phi^K$.

\begin{figure}
\begin{center}
\includegraphics[width=.33\textwidth]{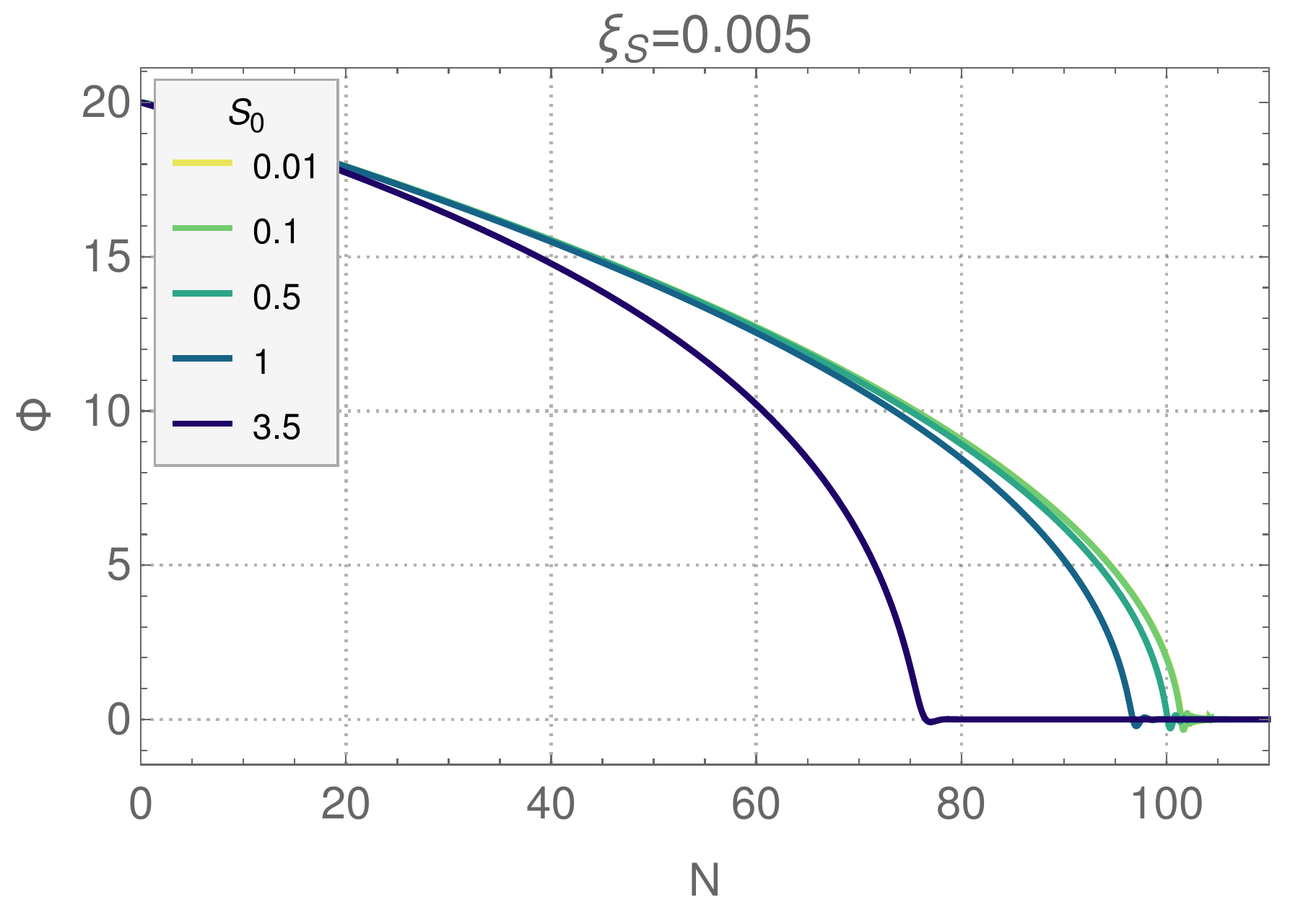}\includegraphics[width=.33\textwidth]{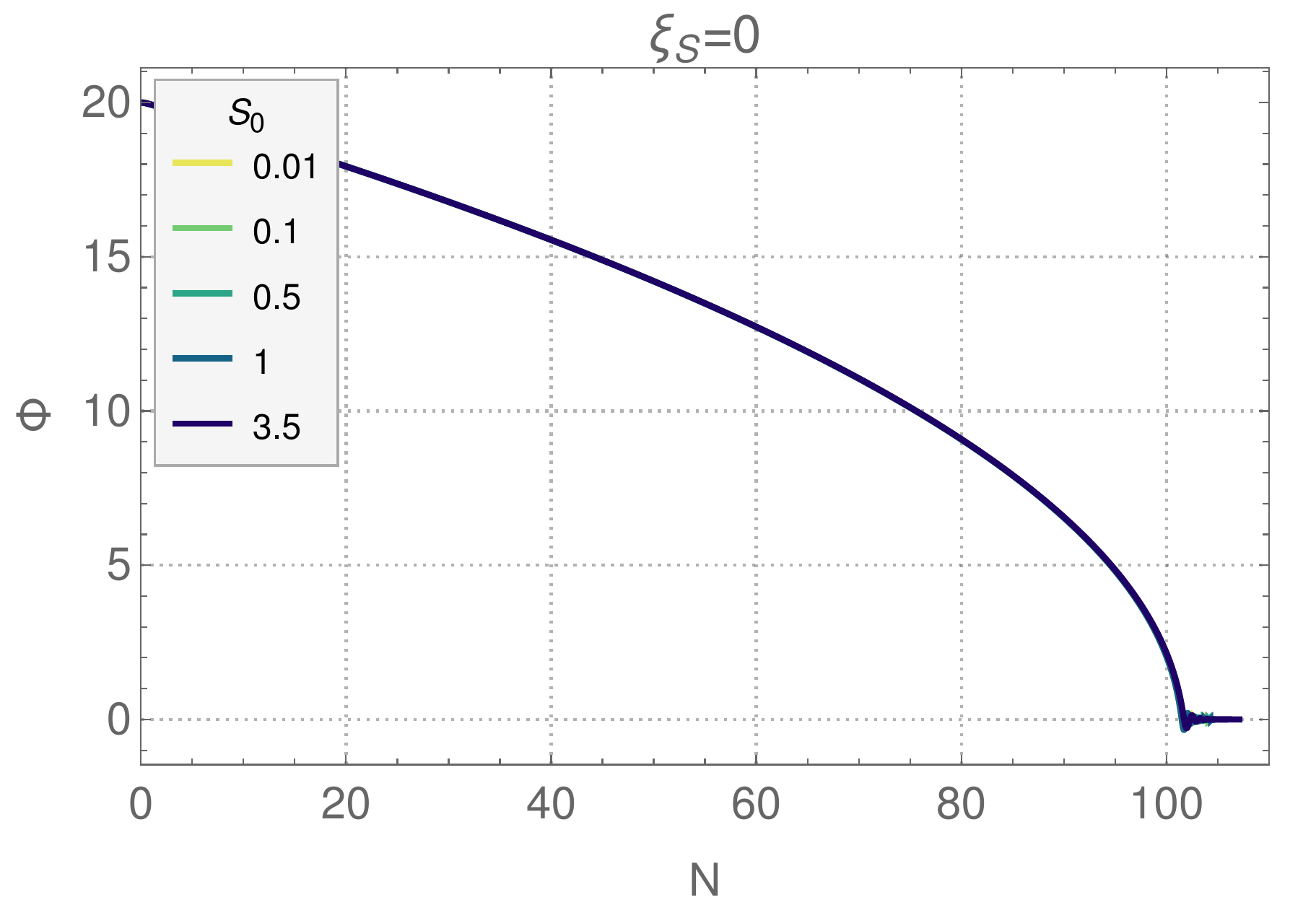}
\includegraphics[width=.33\textwidth]{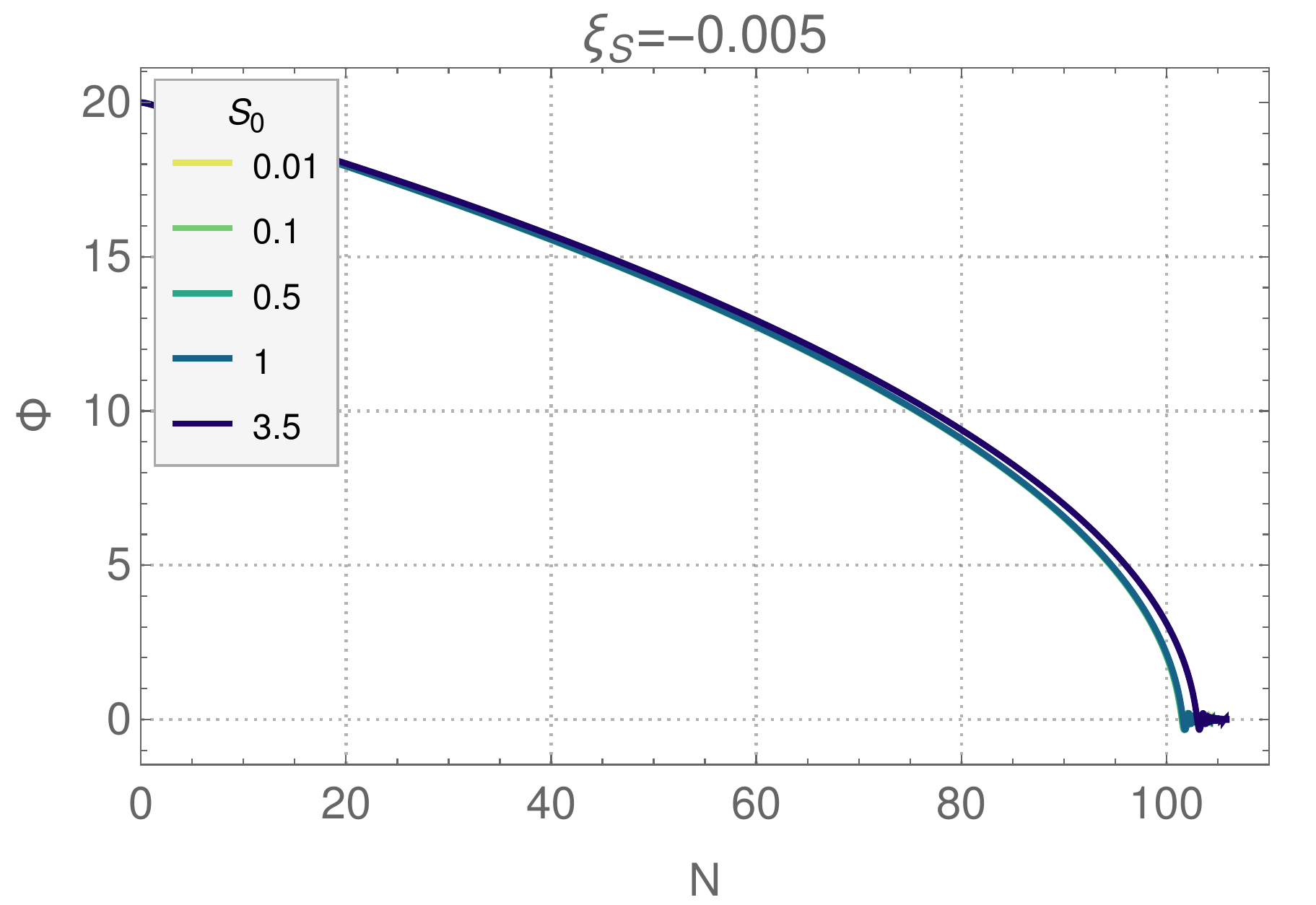}
\\
\includegraphics[width=.33\textwidth]{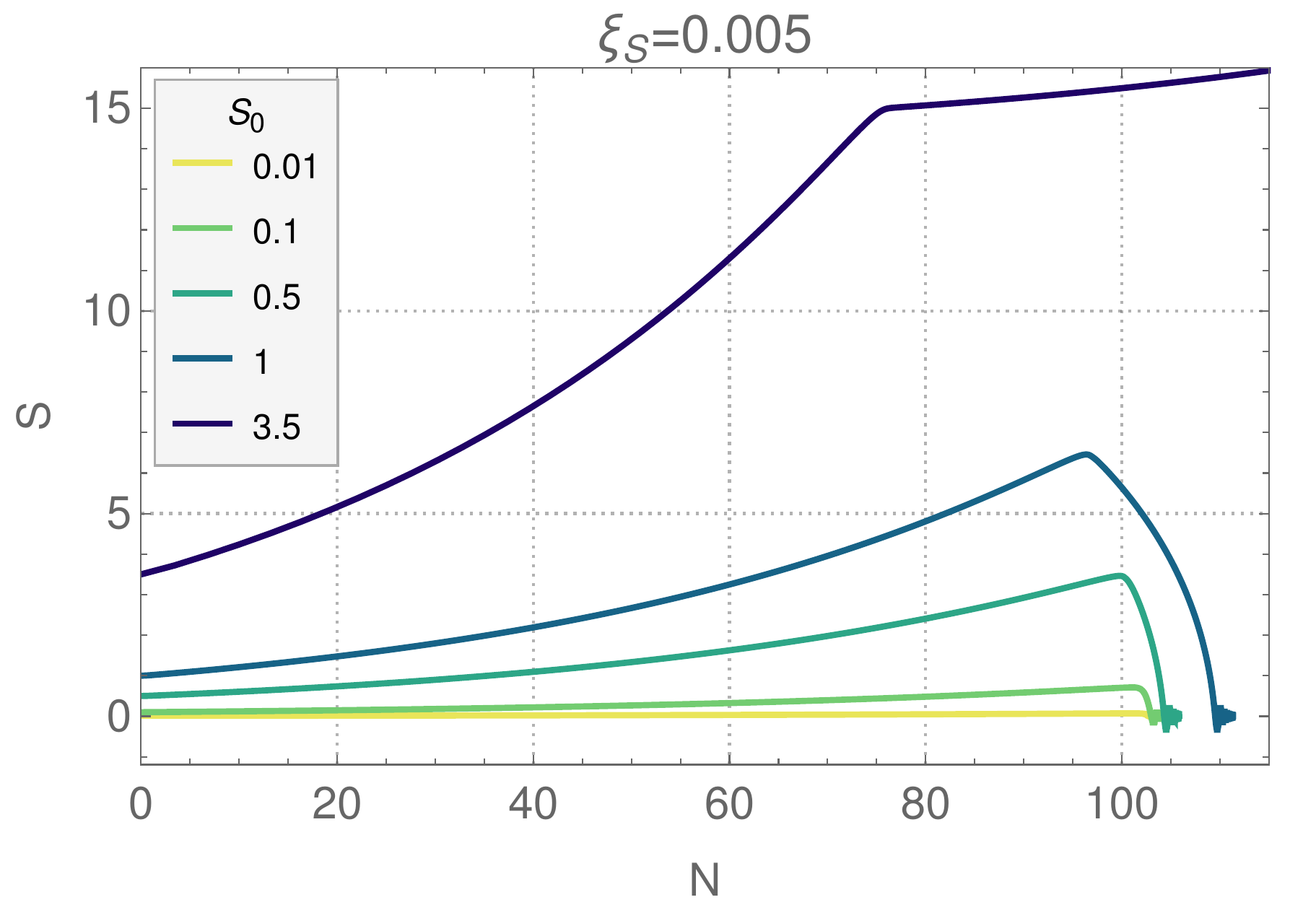}\includegraphics[width=.33\textwidth]{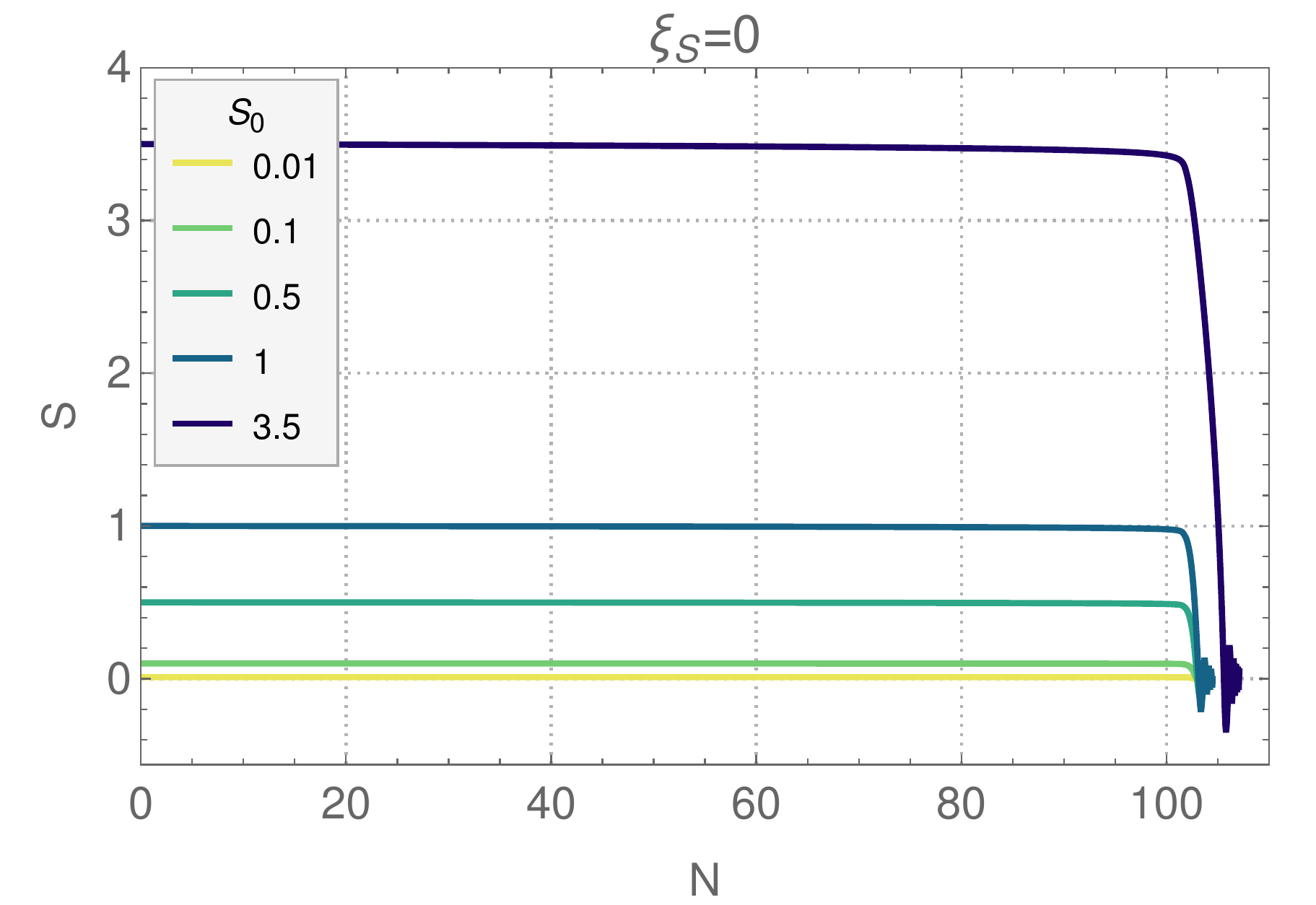}
\includegraphics[width=.33\textwidth]{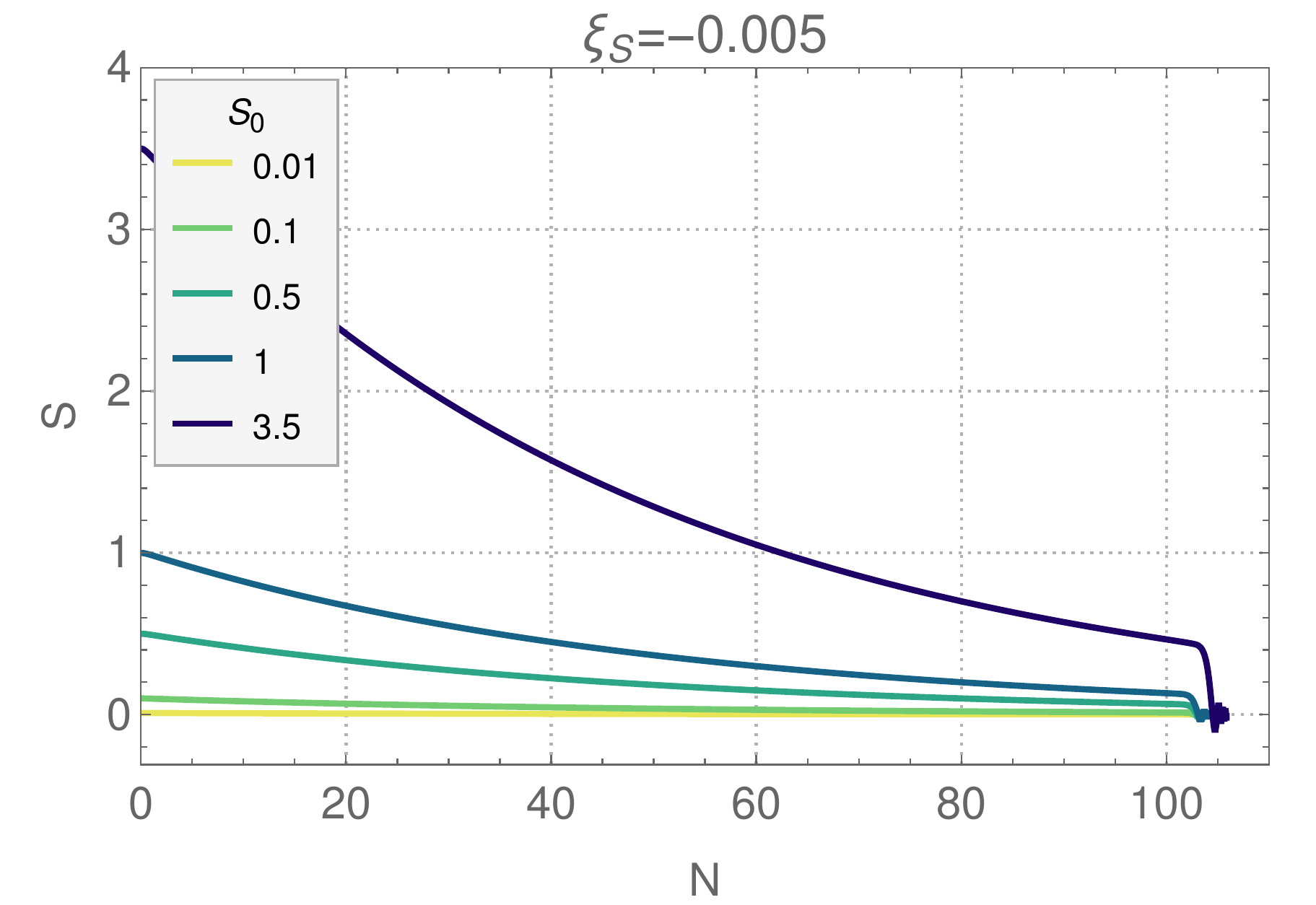}
\caption{Evolution of the initial spectator $S$ for the cases $\xi_S = 0.005$ (left), $0$~(middle), and $-0.005$ (right) as a function of the number of $e$-folds $N$, which is taken to be 0 at the start of our numerical calculation.
The values of $S_0$ in the legend are given in units of $M_{\rm pl}$. \label{fig:evolution_S_BG}}
\end{center}
\end{figure}

In the lower panels of Fig.~\ref{fig:evolution_S_BG}, we show the evolution of the initial spectator $S$ as a function of the number of $e$-folds $N$ for $\xi_s = 0.005$ (left). For reference, we also plot the cases  $\xi_S=0$ (middle) and $\xi_S=-0.005$ (right). The initial value of $\Phi$ is set to be $\Phi_0 = 20 M_{\rm pl}$ and in this setup, the energy density of $\Phi$ dominates over that of $S$ at the beginning of the simulation. The initial inflaton amplitude is not important for the second inflationary stage, as long as it ensures initial domination of the $\Phi$ field because at first the dynamics is effectively single-field and quickly approaches an attractor. The masses of $\Phi$ and $S$ are taken to be $m_\Phi = 5 \times 10^{-6} M_{\rm pl}$ and $m_S = 5 \times 10^{-7} M_{\rm pl}$. The ratio of the masses determines the dynamics, and changing the masses while keeping the ratio fixed merely changes the time scale. In the minimally coupled case $\xi_S=0$ (middle), $S$ slowly rolls down the potential, and when $H \sim m_S$ is reached, it starts to oscillate.  However, when the non-minimal coupling $\xi_S$ is negatively non-zero (right), $S$ decreases faster than the minimally coupled counterpart and becomes more suppressed. On the other hand, when the non-minimal coupling $\xi_S$ is positively non-zero (left), $S$ grows during the initial inflationary period.

\begin{figure}
\begin{center}
\includegraphics[width=8.cm]{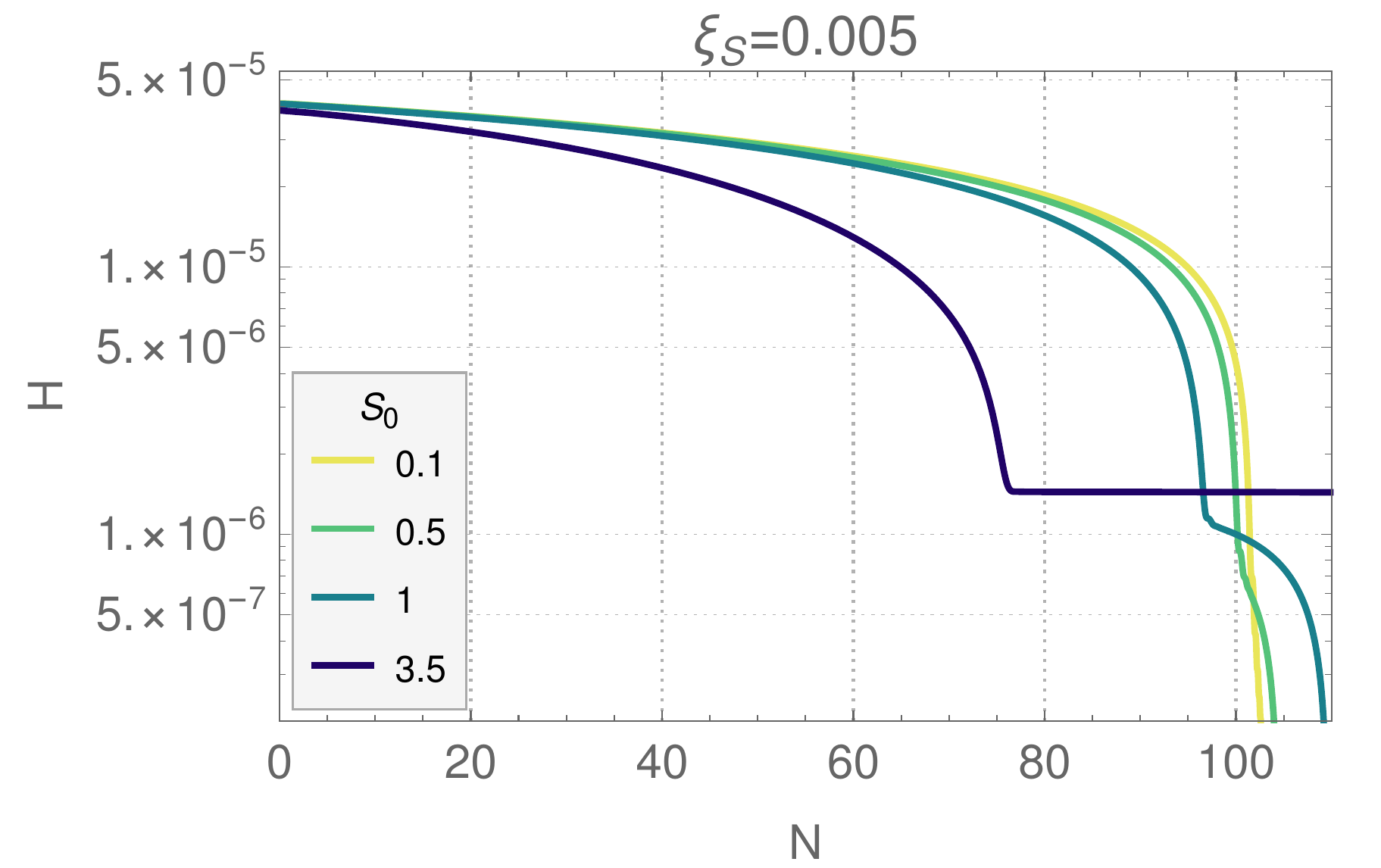}
\caption{Evolution of  $H$ as a function of the number of $e$-folds $N$ for the case of $\xi_S = 0.005$ with several values of $S_0$. 
The values of $S_0$ given in the legend are  in units of $M_{\rm pl}$.  \label{fig:evolution_H_BG}}
\end{center}
\end{figure}

Even if the initial value of $S$ is smaller than $M_{\rm pl}$, $S$ gets increased to super-Planckian field values due to the existence of the spectator non-minimal coupling. Indeed, due to this behavior, the $S$ field can dominate the Universe after the end of the first inflationary phase while it is still rolling slowly, and can drive the second inflationary phase.  This can be confirmed by the evolution of the Hubble parameter in Fig.~\ref{fig:evolution_H_BG}.  As mentioned above, $\Phi$ is initially the dominant component and drives the first period of inflationary expansion. This corresponds to the early phase where the Hubble parameter $H$ in Fig.~\ref{fig:evolution_H_BG} almost stays constant.  Once inflation ends, $H$ sharply drops, which happens at $N \sim 90\text{--}100$ in the figure. However, for the cases of $S_0= M_{\rm pl}$ and $S_0=0.5\Mpl$, $H$ flattens out again, indicating that the second inflationary stage occurs. This second inflation is indeed caused by the initial growth of $S$ seen in the lower-left panel of Fig.~\ref{fig:evolution_S_BG}.

\begin{figure}
\begin{center}
\includegraphics[width=8.cm]{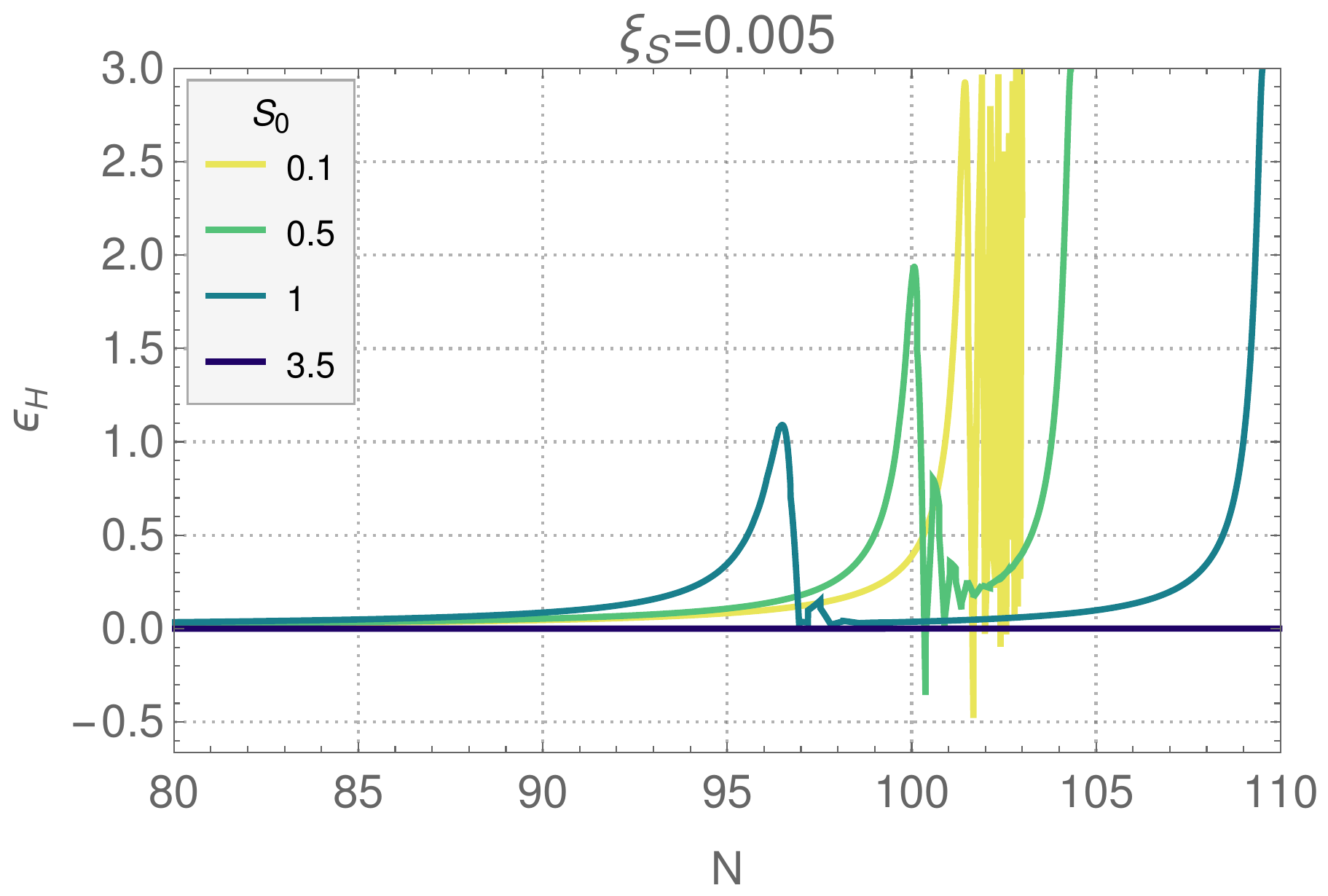}
\caption{Evolution of  $\epsilon_H$ as a function of the number of $e$-folds $N$ for the case of $\xi_S = 0.005$ with several values of $S_0$. 
The values of $S_0$ given in the legend are in units of $M_{\rm pl}$.
\label{fig:evolution_eps_BG}}
\end{center}
\end{figure}

Fig.~\ref{fig:evolution_eps_BG} shows the evolution of the slow-roll parameter $\epsilon_H\equiv -\dot H/H^2$. As can be seen from the figure, $\epsilon_H$ grows steadily during the $\Phi$-driven stage until it exceeds unity and then decreases again as the second inflation begins to be driven by the amplified $S$ field. However, if the initial amplitude is not sufficiently large, as in the case of $S_0=0.1\Mpl$, the field does not have enough time in the first inflationary stage to become dominant enough to drive the second inflationary stage. On the other hand, if the initial value is too large, it will be amplified beyond the point of no return, as we saw from the analytic discussion. We have confirmed this behavior numerically.

\begin{figure}
\begin{center}
\includegraphics[width=9cm]{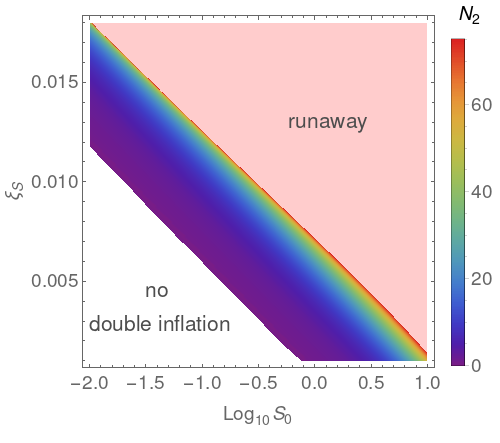}
\caption{The number of $e$-folds generated during the second inflation $N_2$ for mass ratio $m_\Phi / m_S = 10$. In the lower left region (white), there is no second inflation generated by the $S$ field as it never becomes dominant. In the upper right region (salmon), $S$ is amplified past the point of no return.
}
\label{fig:Ne_2nd_inf}
\end{center}
\end{figure}

As a result of this dynamics, there is a range of initial spectator amplitude and non-minimal coupling values where the second inflationary stage can be realized. We have performed a numerical scan of the parameter space to determine the viable region which is shown in Fig.~\ref{fig:Ne_2nd_inf}.\footnote{
The mass ratio in the numerical simulation was taken to be $m_\Phi/m_S=10$. As mentioned earlier, the dynamics of the second inflationary stage is insensitive to the initial inflaton amplitude and absolute mass values.
}
As seen from the figure, even when the initial amplitude of $S$ is sub-Planckian, the second inflation can be generated, and its $e$-folding number $N_2$ can also be very large, $N_2 \gtrsim 60$.\footnote{
In fact, it can be tuned to be arbitrarily large on the edge from the runaway as we can choose initial values such that the field ends up balancing on the ridge of the Einstein frame potential.
}
If the field overshoots and ends up in the runaway phase, the second phase of inflation never terminates.  Such cases correspond to the upper right region in the figure. The lower white corner corresponds to cases where the $S$ field never becomes dominant.

\section{Evolution of fluctuations and power spectra \label{sec:perturbation}}
In this section, we discuss the evolution of fluctuations in our setup. First, we briefly summarize how we calculate the perturbations in the model. We use the formalism of \cite{Langlois:2008mn,Kaiser:2010yu,Kaiser:2012ak} for decomposing the fluctuations into adiabatic and entropic components as follows.  A scalar field $\varphi^I$ can be divided into the background part $\bar{\varphi}^I$ and its perturbation $Q^I$ as
\begin{equation}
\varphi^I = \bar{\varphi}^I + Q^I \,.
\end{equation}
We can define a new basis $\{e_n^I\} ~(n=\sigma, {s})$ such that the first vector points along the background trajectory, corresponding to the adiabatic mode whose basis vector can be written as
\begin{equation}
 e_\sigma^I \equiv  \frac{\dot\varphi^I}{\sqrt{G_{JK}\dot\varphi^J\dot\varphi^K}} = \frac{\dot\varphi^I}{\dotsig} \,,
\end{equation}
where we have defined $\dotsig \equiv \sqrt{G_{IJ}\dot\varphi^I\dot\varphi^J}$. The remaining isocurvature direction is then perpendicular to the background trajectory. The perturbations $Q^I$ in the new basis are written as $Q^I = Q^ne_n^I$ with $n=\sigma, s$.

The equations of motion for the adiabatic and isocurvature modes $Q_\sigma$ and $Q_{s}$ are
\begin{align}
 \ddot{Q}_{\sigma} + 3H \dot{Q}_{\sigma} + \left(\frac{k^2}{a^2} + \mu_{\sigma}^2\right)Q_{\sigma} 
 & = 
 \frac{\ud}{\ud t}\Big(\Xi Q_{s} \Big) - \left(\frac{\dot H}{H} + \frac{V_{,\sigma}}{\dotsig}\right)\Xi Q_{s}    
 \label{eq:eom_Q_sigma}
\intertext{and}
\label{eq:eom_Q_s}
 \ddot{Q}_{s}  +  3H \dot{Q}_{s} + \left(\frac{k^2}{a^2} + \mu_{s}^2\right)Q_{s} 
 &= -\Xi\left[\dot{Q}_{\sigma} + \left(\frac{\dot{H}}{H} - \frac{\ddotsig}{\dotsig}\right)Q_{\sigma}\right]  ,
\end{align}
respectively, where $\mu_\sigma$, $\mu_{s}$, and $\Xi$ are given by 
\begin{eqnarray}
 \mu_{\sigma}^2 & = & V_{;\sigma \sigma} - \frac{1}{a^3}\frac{\ud}{\ud t}\left(\frac{a^3\dotsig^2}{H}\right) -  \frac{V_{,{s}}^2}{\dotsig^2}, \label{eq:mu_sigma}
 \\
 \mu_{s}^2 & = &   V_{; {ss}} + \frac{1}{2}\dotsig^2\tilde{\mathcal{R}}  - \frac{V_{, {s}}^2}{\dotsig^2} , \label{eq:mu_s}
 \\
 \Xi & = & -\frac{2V_{,{s}}}{\dotsig }  \,,
  \label{eq:Xi}
\end{eqnarray}
in which $V_{;\sigma \sigma}$ and $V_{;{ss}}$ are defined as projections of $\mathcal{D}_I\mathcal{D}_JV$ onto the appropriate directions and $\tilde{\mathcal{R}}$ is  the Ricci curvature scalar of the field manifold defined by $G_{IJ}$.  The comoving curvature perturbation in the flat gauge  is given as
\begin{equation}
  \mathcal{R} = \frac{H}{\dotsig}Q_{\sigma} \,. 
\end{equation}
The evolution of ${\cal R}$ can be calculated by solving the equations of motion \eqref{eq:eom_Q_sigma} and \eqref{eq:eom_Q_s} along with the background ones. The primordial (scalar) power spectrum ${\cal P_R}$ is computed by  evaluating ${\cal R}$ at the end of inflation and given by 
\begin{eqnarray}
{\cal P}_\zeta (k) &=&  \frac{k^3}{2\pi^2} |{\cal R_\mathbf{k}}|^2 \,,
\end{eqnarray}
while its scale dependence is characterized by the spectral index
\begin{equation}
\label{eq:ns_def}
n_s - 1 \equiv \frac{ d \log {\cal P_\zeta}}{d \log k}  \,. 
\end{equation}

We also discuss primordial gravitational waves generated during inflation. Gravitational waves correspond to the tensor perturbations whose power spectrum can be calculated with the standard procedure (see, e.g., \cite{Lidsey:1995np}) as
\begin{equation}
{\cal P}_T (k) =  \frac{k^3}{2\pi^2} |v_k|^2 \,,
\end{equation}
where $v_k$ is the Fourier component of $v_{ij} = M_{\rm pl} h_{ij} /(2a)$  in a plane wave expansion, with $h_{ij}$ being the tensor perturbation in the FRW metric. $v_k$ satisfies the following equation of motion:
\begin{equation}
\ddot{v_k} + H \dot{v_k} + \left( \frac{k^2}{a^2} - H^2 - \frac{\ddot{a}}{a} \right) v_k = 0.
	\label{tensor eom}
\end{equation}
To quantify the size of the tensor power spectrum, the tensor-to-scalar ratio $r$ is commonly adopted, which can be compared with observational constraints.  $r$ is defined at some reference scale $k=k_\ast$ as 
\begin{equation}
\label{eq:def_r}
r \equiv \left. \frac{{\cal P}_T}{{\cal P}_\zeta} \right|_{k =k_\ast} \,.
\end{equation}

\subsection{Numerical analysis}

In our numerical analysis, we solve the fluctuation equations of motion~\eqref{eq:eom_Q_sigma}, \eqref{eq:eom_Q_s}, and \eqref{tensor eom} altogether. We evolve the fluctuations from an initial time inside the horizon, through the first inflationary stage, to the end of the second inflation.  We identify the reference scale $k_\ast$ that has exited the horizon 60 $e$-folds before the end of the second stage of inflation. We normalize the masses to match the observed spectrum amplitude $\mathcal P_{\mathcal R} \simeq 2.2\times 10^{-9}$ at the reference scale, with the mass ratio being fixed.

Following~\cite{Lalak:2007vi,vandeBruck:2014ata,Braglia:2020fms}, we solve the evolution of fluctuations twice: once with $Q_s = 0$ and the initial Bunch-Davies vacuum amplitude $Q_\sigma = 1/\sqrt{2k}$ and once more with the opposite choice. This ensures the absence of spurious cross-correlations in power spectra. We have also solved the evolution of fluctuations in the original basis $(\Phi, S)$ and verified that the results agree as long as this technique is employed.

\subsection{Scalar and tensor power spectra}   

In this section, we discuss the predictions for scalar and tensor power spectra in the model. To obtain the spectra we scan through the momentum space, solving the evolution of fluctuations as outlined above for each $k$-mode. We focus specifically on the case where the second inflation is driven by the $S$ field which has grown to reach a super-Planckian value due to the existence of the non-minimal coupling.\footnote{
For numerical studies of power spectra in double inflation with two quadratic potentials in the minimally coupled case, see e.g. \cite{Feng:2003zua,Braglia:2020fms}.
}

\begin{figure}[t]
\begin{center}
\includegraphics[width=.48\textwidth]{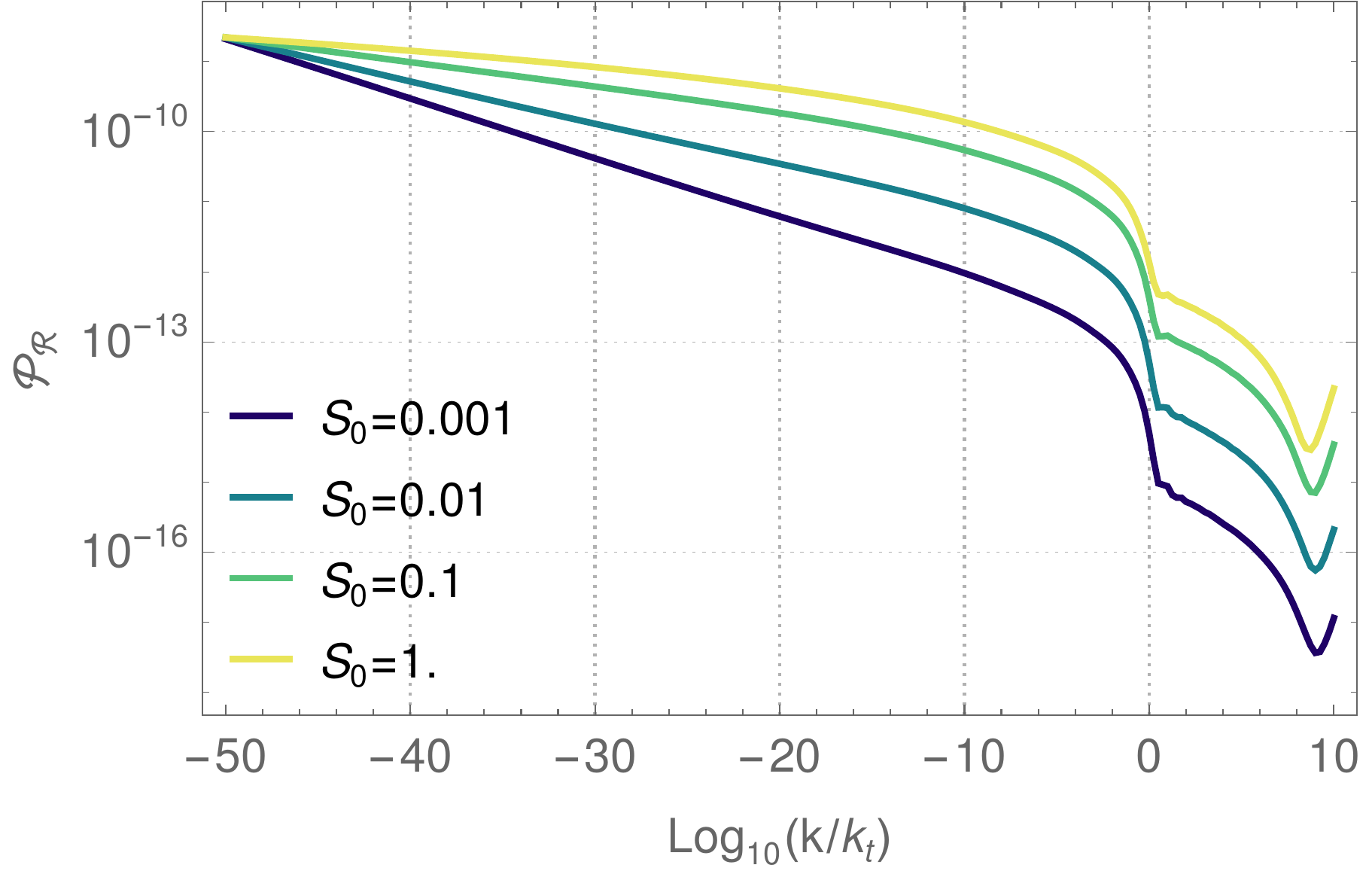}\includegraphics[width=.48\textwidth]{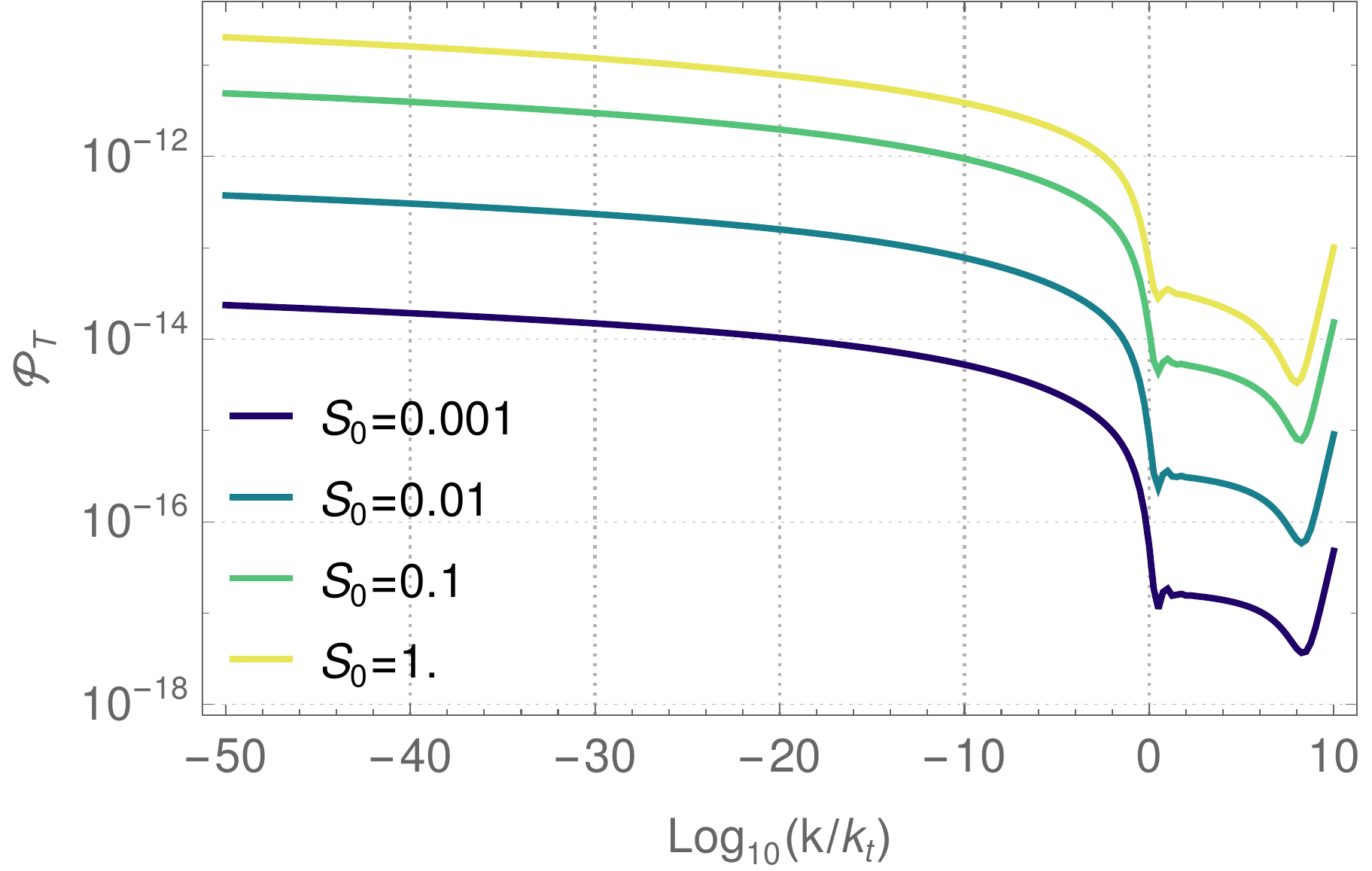}
\caption{\label{fig:power_spectrum_1}
Scalar and tensor power spectra (right and left panels, respectively) for several values of $S_0$. $\xi_S$ is fixed by requiring that the second inflation lasts for $N_2=10$, which gives $\xi_S =  0.005$, 0.011, 0.016, and 0.022 for $S_0= M_{\rm pl}$, $10^{-1}M_{\rm pl}$, $10^{-2}M_{\rm pl}$, and $10^{-3}M_{\rm pl}$, respectively. The ratio of the masses is taken to be $m_\Phi / m_S = 10$. The value of $m_\Phi$ (or $m_S$) is fixed by requiring that the scalar power spectrum at $\log (k_\ast/k_t) =-50$ is normalized to ${\cal P}_{\cal R} = 2.2 \times 10^{-9}$. 
}
\end{center}
\end{figure}

In Fig.~\ref{fig:power_spectrum_1},  we show scalar and tensor power spectra for several values of $S_0$. The value of $\xi_S$ is chosen such that the second inflationary stage lasts $N_2=10$ $e$-folds for each $S_0$. The resultant values are respectively $\xi_S =  0.005$, 0.011, 0.016, and 0.022 for $S_0= M_{\rm pl}$, $10^{-1}M_{\rm pl}$, $10^{-2}M_{\rm pl}$, and $10^{-3}M_{\rm pl}$. We have fixed the mass ratio to be $m_\Phi / m_S = 10$.  The power spectra are plotted as a function of $\log (k/ k_t)$, where $k_t$ is the wave number of the mode that exited the horizon at the beginning of the second inflation.
The size of $m_\Phi$ is fixed by requiring that the scalar power spectrum at the reference scale $k_\ast$, given as $\log (k_\ast/k_t) =-50$ in the current choice, is normalized to ${\cal P}_{\cal R} = 2.2 \times 10^{-9}$.

Both scalar and tensor power spectra are suppressed for the modes $k > k_t$ having exited the horizon during the second inflation. This is mainly because the first inflation, during which the modes $k < k_t$ exited the horizon, is driven by $\Phi$ whose mass is assumed to be larger than that of $S$. Although the shape of the power spectra generally varies depending on the value of $S_0$ (and $\xi_S$), the main overall feature remains the same regardless of the values of the parameters. 

Regarding the behavior of the tensor power spectrum ${\cal P}_T$, effects of the spectator non-minimal coupling $\xi_S$ can be understood by noticing that, when $\xi_S$ gets larger (or $S_0$ gets smaller to keep $N_2=10$), the adiabatic fluctuations are more sourced by the isocurvature ones, in which the intrinsic adiabatic mode should give less contribution to the total one and the mass of $\Phi$ is lowered to satisfy the normalization condition for ${\cal P_R}$.  This results in the decrease of the Hubble parameter during inflation. Since the tensor power spectrum is given as ${\cal P}_T \propto (H_{\rm inf} / M_{\rm pl})^2$ with $H_{\rm inf}$ being the Hubble parameter during inflation, the amplitude of the tensor power spectrum is lowered as $S_0$ decreases.  Furthermore, sourcing of isocurvature fluctuations is more prominent on large scales since the sourcing continues longer for the modes which exited the horizon earlier. 
Since we fix the number of the second inflation to be $N_2=10$, the case with smaller $S_0$ experiences more sourcing of isocurvature fluctuation,  and thus ${\cal P_R}$ for smaller $S_0$ gets more red-tilted.

\begin{figure}[t]
\begin{center}
\includegraphics[width=.48\textwidth]{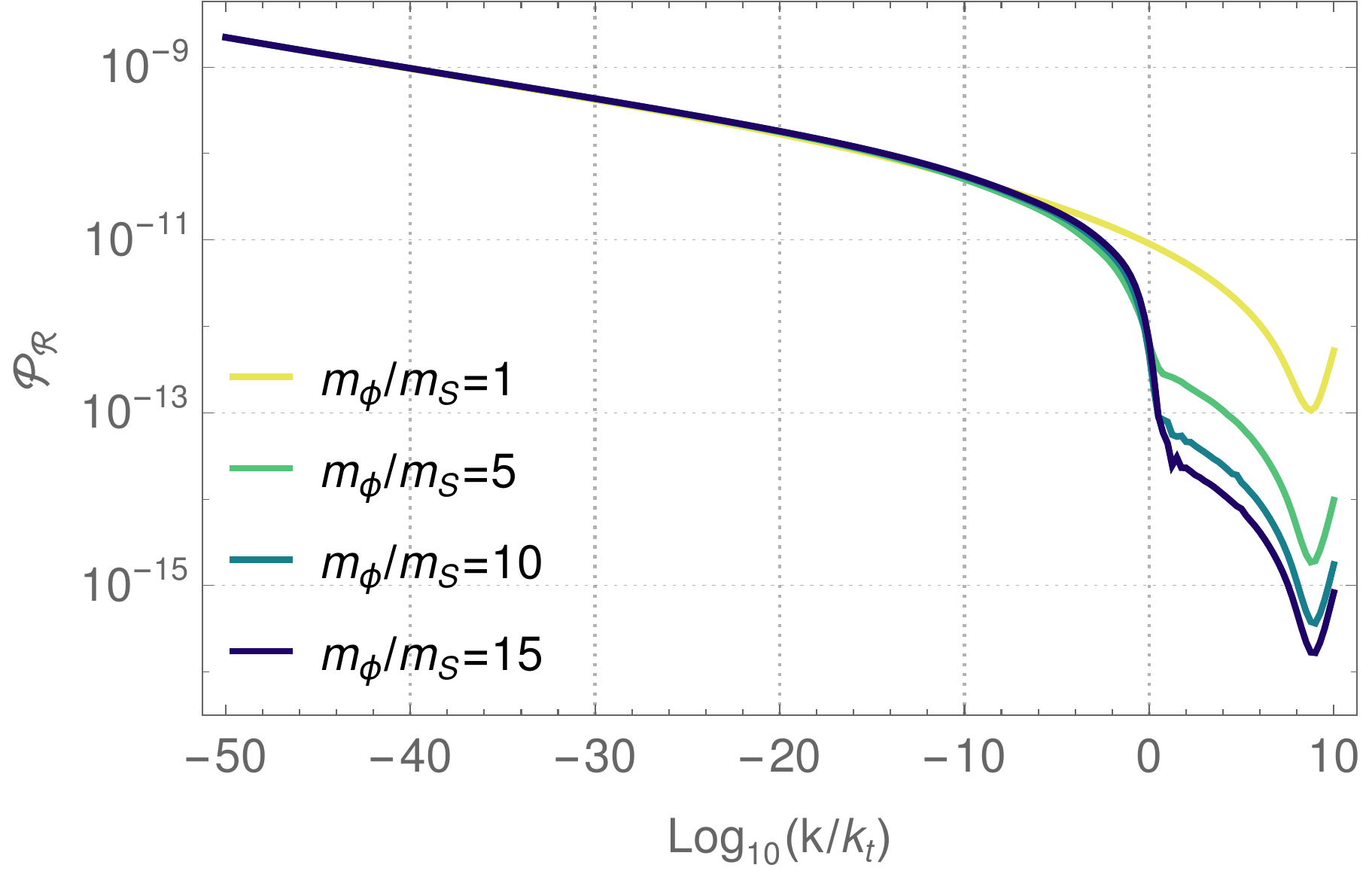}\includegraphics[width=.48\textwidth]{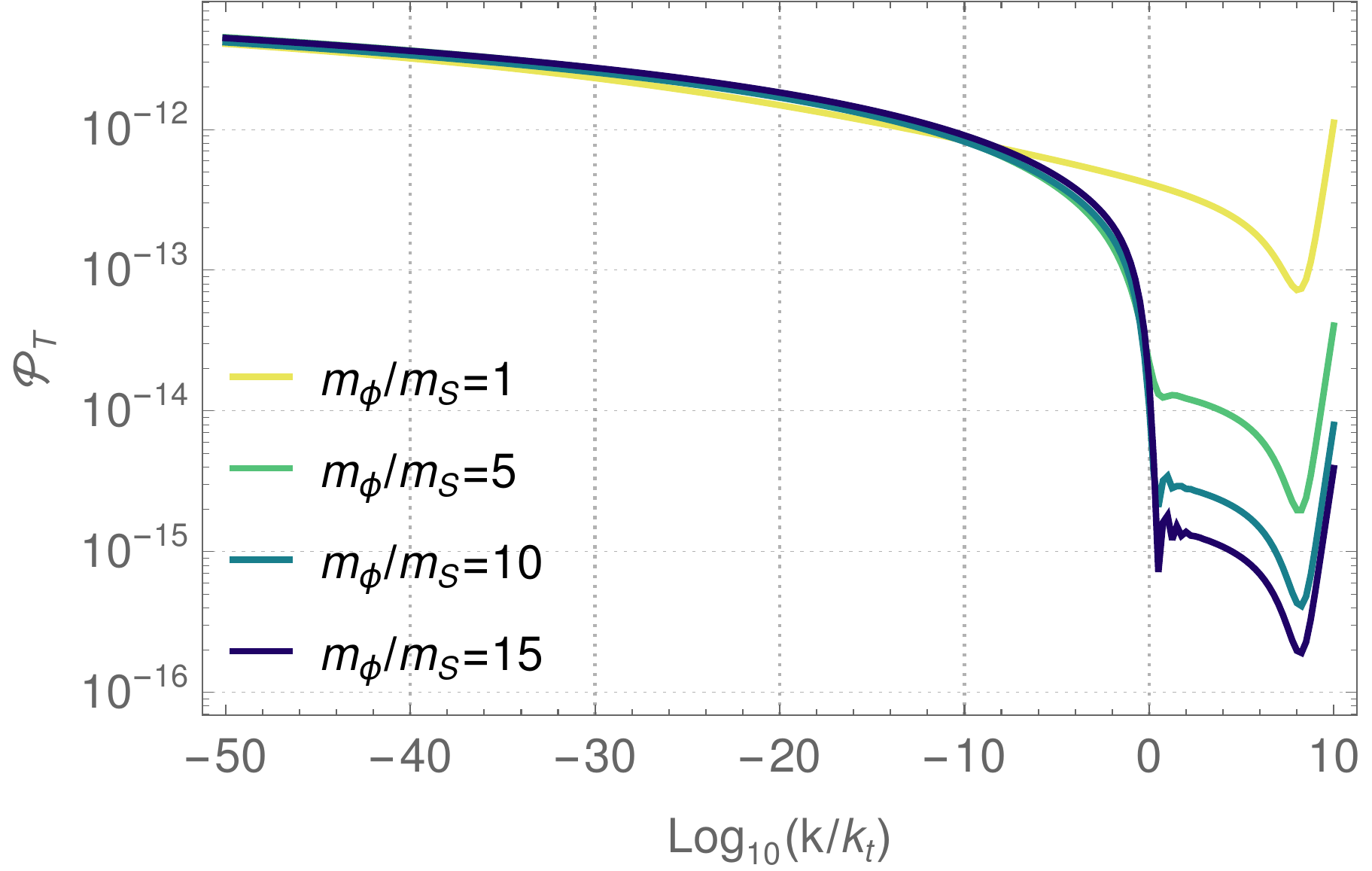}
\caption{\label{fig:power_spectrum_2}
Scalar and tensor power spectra (right and left panels, respectively) for several values of the ratio $m_\Phi/m_S =1$, 5, 10, and 15. We take $S_0 = 0.1 M_{\rm pl}$ and $\xi_S \simeq 0.01$, such that the number of $e$-folds during the second inflation is $N_2=10$.  The value of $m_\Phi$ is fixed by requiring that the scalar power spectrum is normalized at $\log (k_\ast/k_t) =-50$ as $P_\zeta = 2.2 \times 10^{-9}$.  
}
\end{center}
\end{figure}

In Fig.~\ref{fig:power_spectrum_2}, scalar and tensor power spectra for several values of  the ratio $m_\Phi / m_S$ are shown. The values of $S_0$ is fixed at $S_0 =  0.1 M_{\rm pl}$ and $\xi_S$ is once again fixed by requiring that the number of $e$-folds during the second inflation is $N_2=10$. However, in practice, $\xi_S \simeq 0.01$ for all cases shown.  As can be seen from the figure, the ratio $m_\Phi / m_S$ mainly affects the power spectra for modes with $k > k_t$ which exited the horizon in the second inflation driven by $S$.   Since $m_\Phi$ is effectively fixed by the normalization to the observed spectrum, increasing the ratio amounts to decreasing $m_S$. Therefore, for the modes $k > k_t$, the situation is the same as the one for a single-field quadratic chaotic inflation without a non-minimal coupling, and lowering $m_S$  means that the scalar power spectrum is suppressed. The tensor mode also gets decreased since the Hubble rate during the second inflation is lowered.  Although both power spectra ${\cal P_R}$ and ${\cal P}_T$ seem to be almost unchanged for $k < k_t$, when the mass ratio is decreased to be close to unity,  adiabatic perturbations are more sourced by isocurvature ones, which indicates that ${\cal P_R}$ gets more red-tilted.  Furthermore, this means that the tensor-to-scalar ratio is suppressed since only the scalar mode is enhanced. This trend is more clearly seen in terms of the spectral index and the tensor-to-scalar ratio as we discuss below.

\subsection{Spectral index and tensor to scalar ratio}

We calculate $n_s$ and $r$ at the reference scale corresponding to the mode which exited the horizon at $ \log (k/k_t) = -50$.  Since we calculate the power spectrum ${\cal P}_{\cal R}$ numerically, the spectral index, defined in Eq.~\eqref{eq:ns_def},  is obtained by numerically evaluating the derivative of ${\cal P}_{\cal R}$ with respect to $k$. For the tensor-to-scalar ratio $r$, defined in Eq.~\eqref{eq:def_r}, we take the ratio of  ${\cal P_R}$ and ${\cal P}_T$ at the above mentioned reference scale. 

\begin{figure}
\begin{center}
\includegraphics[width=8cm]{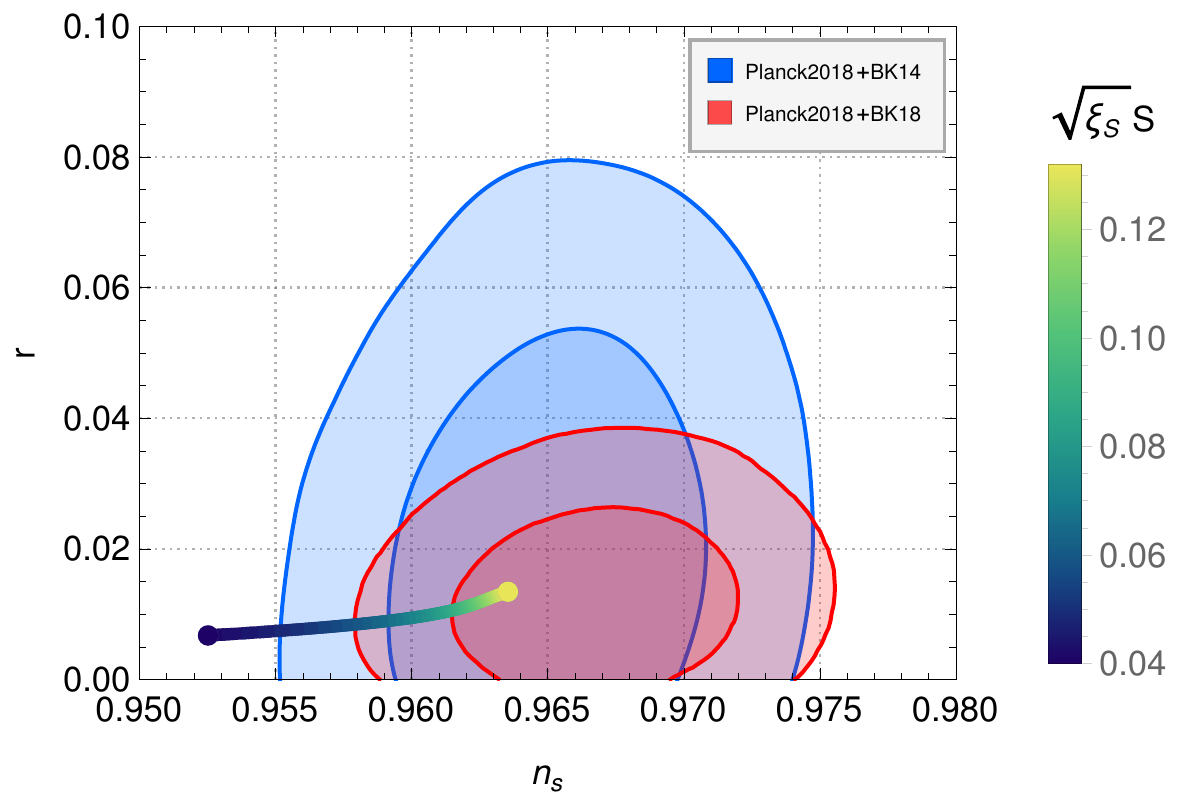} 
\caption{ \label{fig:ns_r_xi} 
Spectral index $n_s$~vs.~the tensor-to-scalar ratio $r$ predicted in the model varying $\sqrt{\xi_S}S_0$. The values of $\xi_S$ and $S_0$ are chosen to keep $N_2=10$ unchanged. The mass ratio is fixed as $m_\Phi/m_S =10$. Observational constraints on $n_s$ and $r$ from Planck~\cite{Planck:2018jri} and Planck+BICEP/Keck~2018~\cite{BICEP:2021xfz} are also depicted. 
}
\end{center}
\end{figure}

In Fig.~\ref{fig:ns_r_xi}, the predictions for $n_s$ and $r$ varying $\sqrt{\xi_S} S_0$ are shown. Since the combination of $\xi_S S^2$ appears in the function $f(S)$ which represents a non-minimal coupling in the action, this is the main measure of the effect, though the additional difference can be introduced by the kinetic mixing. In practice, we vary the value of $S_0$ and fix $\xi_S$ to keep the duration of the second inflation at $N_2=10$. As already discussed earlier, as $S_0$ (or $\sqrt{\xi_S} S_0$) decreases, ${\cal P_R}$ gets red-tilted and ${\cal P}_T$ is suppressed, which explains the behavior of the predictions in the $n_s$--$r$ plane.  Notice that the limit of $\sqrt{\xi_S}S_0 \rightarrow 0$ does not correspond to a minimally coupled case because here we fix the number of $e$-folds during the second inflation to be $N_2=10$. In this case, even when $\sqrt{\xi_S}S_0$ is assumed to be small, we need to take a large value of $\xi_S$ even for a small value of $S_0$, which is the reason why $n_s$ gets more red-tilted for a smaller $\sqrt{\xi_S} S_0$.

\begin{figure}
\begin{center}
\includegraphics[width=8cm]{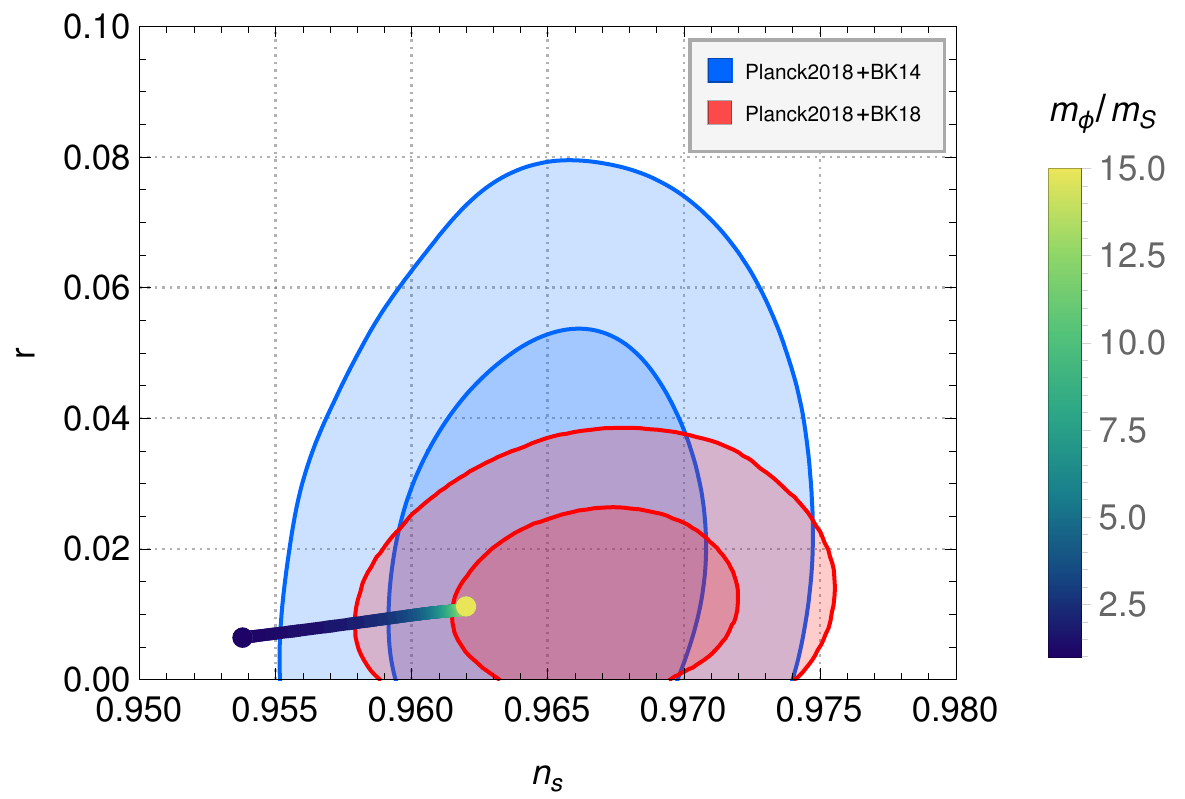}
\caption{\label{fig:ns_r_mratio}
Dependence of $n_S$ and $r$ on the mass ratio $m_\Phi/m_S$ with $S_0=1.5 M_{\rm pl}$. $\xi_S$ is chosen such that the duration of the second inflation is $N_2=10$. 
}
\end{center}
\end{figure}

The dependence of $n_s$ and $r$ on the mass ratio $m_\Phi / m_S$ is shown in Fig.~\ref{fig:ns_r_mratio}. The value of $S_0$  is fixed as $S_0 = 1.5 M_{\rm pl}$ and the number of $e$-folds during the second inflation is again taken to be 10, which determines the value of $\xi_S$. When the ratio is $m_\Phi / m_S \gtrsim 7 $, $\xi_S$ is almost unchanged and takes the value $\xi_S \simeq 0.0036$.  As already mentioned, when $m_\Phi / m_S$ is decreased to be close to unity, adiabatic perturbations are more sourced by the isocurvature ones, which induces more scale-dependence and thus the spectral index becomes smaller. When the sourcing from isocurvature fluctuations is stronger, ${\cal P_R}$ is enhanced, and the tensor-to-scalar ratio decreases.  Note that larger values of $m_\Phi / m_S$ are favorable since as the ratio increases, the predictions in the $n_s$--$r$ plane move to the center of the current cosmological constraints. Although the tensor-to-scalar ratio is generically suppressed in our setup, the scalar power spectrum becomes more red-tilted when more adiabatic fluctuations are sourced by the isocurvature mode, and thus the mass ratio should be taken not to be inconsistent with the measurement of $n_s$.

\begin{figure}
\begin{center}
\includegraphics[width=8cm]{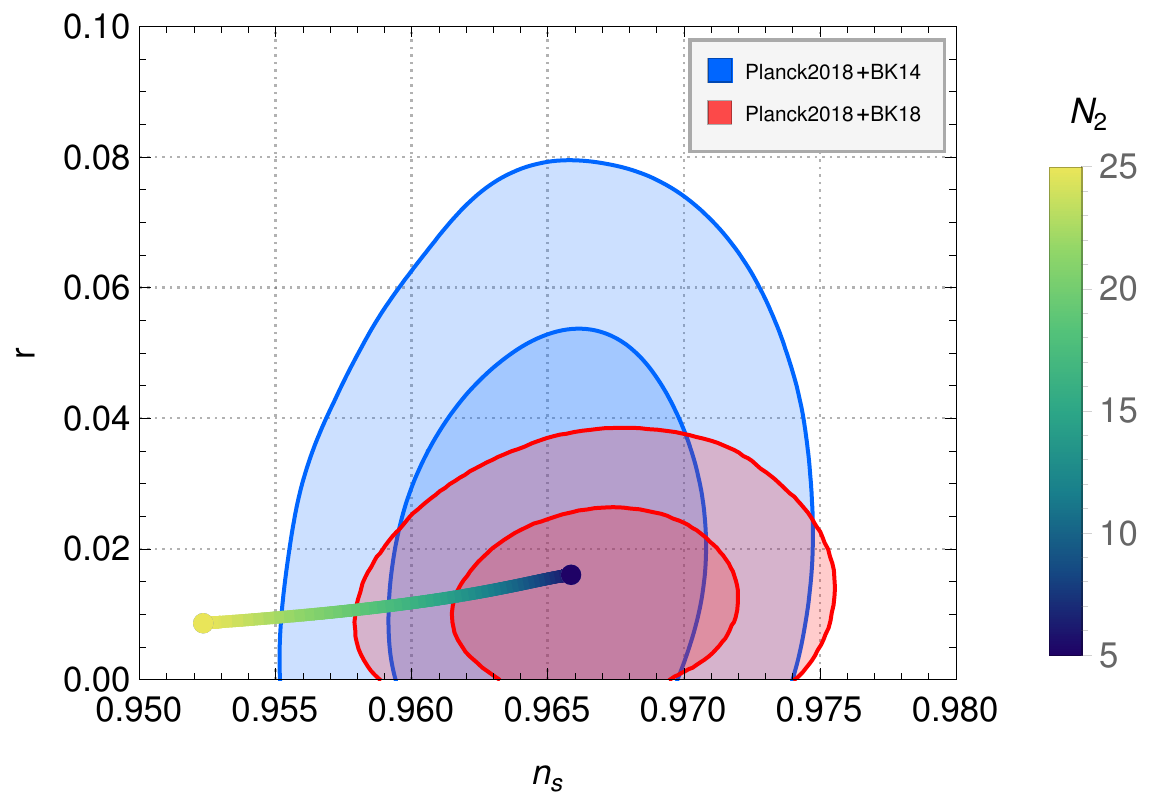}
\caption{\label{fig:ns_r_N2}
Dependence of $n_s$ and $r$ on duration of second inflation $N_2$. The initial value of $S$ is fixed as $S_0=4 M_{\rm pl}$ and $\xi_S$ is chosen such that the value of $N_2$ depicted in the color bar is realized. The mass ratio is taken to be $m_\Phi/m_S=10$.
}
\end{center}
\end{figure}

Finally, we show the predictions of our model in the $n_s$--$r$ plane, varying the number of $e$-folds during the second inflation. In the figure, the mass ratio is fixed to be $m_\Phi /m_S =10$ and the initial value of $S$ to be $S_0 = 4 M_{\rm pl}$. The value of $\xi_S$ is then fixed by requiring a specific value of $N_2$. To have a large $N_2$, we need to have a bigger $\xi_S$ since more growth of $S$ is necessary to have a longer period of the second inflation.  Thus, when $N_2$ is larger, the effects of isocurvature field $S$ becomes bigger (more adiabatic perturbations are sourced by $S$). Therefore large $N_2$ indicates that ${\cal P_R}$ is more red-tilted and $r$ is more suppressed.

In Figs.~\ref{fig:ns_r_xi}, \ref{fig:ns_r_mratio}, and \ref{fig:ns_r_N2},   observational constraints from Planck \cite{Planck:2018jri} as well as from Planck+BICEP/Keck 2018 \cite{BICEP:2021xfz} are superimposed. As can be seen, the spectral index and the tensor-to-scalar ratio are predicted to be well within the current cosmological constraints for a range of model parameters,  indicating that a quadratic chaotic inflation model can be rescued by introducing a spectator field non-minimally coupled to gravity.

\section{Conclusion and Discussion} \label{sec:conclusion}

In this paper, we have investigated the two-field inflation model where the field $S$ that initially plays the role of spectator comes to dominate and causes the second inflation due to the non-minimal coupling to gravity. We have first studied the evolution of the background quantities of the fields and have shown that the second inflationary phase is generically realized due to the existence of the spectator non-minimal coupling $\xi_S$.  When $\xi_S$ is positive, $S$ field can grow during the first inflationary phase driven by $\Phi$ and once $S$ acquires a super-Planckian field value, it can generate the second inflation. However, when $\xi_S$ or the initial amplitude is too large a runaway behavior can result where the field keeps growing and the second inflation does not terminate. However, in a sizable region of parameter space, the $S$ field drives a finite amount of inflation and then returns to zero.

We also investigated the power spectra for both the scalar and tensor modes in the model. As discussed in Sec.~\ref{sec:perturbation}, adiabatic fluctuations are more sourced by isocurvature ones as $\xi_S S^2$ (or $\xi_S$) increases. Due to this isocurvature sourcing augmented by the spectator non-minimal coupling, the scale-dependence of the scalar power spectrum is modified. Although the tensor power spectrum itself is not directly affected by this isocurvature sourcing,  the overall amplitude can change through the requirement of the normalization condition imposed on the scalar power spectrum. But due to the change of the scalar mode, the tensor-to-scalar ratio $r$ can also be affected.  We explicitly showed  $n_s$ and $r$ are modified and can fit well within the current observational constraints. With this setup, the quadratic chaotic inflation model can become viable by assuming a spectator with non-minimal coupling to gravity.  

Multi-field dynamics is common to high energy models and non-minimal coupling to gravity is a generic feature of quantum field theory in curved space-time. Therefore, it is necessary to consider the interplay of such features in the early Universe. As we have shown, even simple dynamics can have a significant impact on cosmological predictions and change the picture considerably from the single-field case.

\section*{Acknowledgements}
This work is supported in part by the Communidad de Madrid “Atracci\'{o}n de Talento investigador” Grant No. 2017-T1/TIC-5305 and the MICINN (Spain) project PID2019-107394GB-I00 (SR) as well as by JSPS KAKENHI Grant Numbers 19H01899 (KO), 21H01107 (KO), 17H01131 (TT), and 19K03874 (TT) and MEXT KAKENHI Grant Number 19H05110 (TT).

\bibliographystyle{JHEP}
\bibliography{refs}

\end{document}